\documentclass[11pt,a4paper]{article}

\usepackage[T1]{fontenc}
\usepackage[utf8]{inputenc}
\usepackage[margin=1in]{geometry}
\usepackage{amsmath,amssymb,amsthm}
\usepackage{booktabs}
\usepackage{graphicx}
\usepackage{xcolor}
\usepackage[authoryear,round]{natbib}
\usepackage[colorlinks=true,allcolors=blue!60!black]{hyperref}
\usepackage{setspace}
\graphicspath{{figures/}}

\newcommand{\Normal}{\mathrm{N}}
\newcommand{\Poisson}{\mathrm{Poisson}}
\newcommand{\Binomial}{\mathrm{Binomial}}
\newcommand{\logit}{\mathrm{logit}}
\newcommand{\alttext}[1]{}

\title{Relegation, promotion and the components of scoring in water polo}

\author{Manuele Leonelli\thanks{School of Science and Technology, IE University,
Madrid, Spain. Email: \href{mailto:manuele.leonelli@ie.edu}{manuele.leonelli@ie.edu}.
ORCID: 0000-0002-2562-5192.}}

\date{\today}

\begin{document}
\maketitle

\begin{abstract}
Relegation, not the championship, decides the outcome of many European water
polo leagues: one club has won every recent national title in Italy, Spain and
Hungary. We make the relegation decision our object of inference. Serie A1
relegates one club for finishing last and a second through a three-match
play-out, and we ask at each stage how much can be known and when. Promoted
clubs arrive with no top-flight record, so we also compare four ways of
setting their prior. These analyses rest on the first
Bayesian hierarchical model of water polo, which splits scoring into
even-strength, man-up and penalty components, models opportunities and
conversions separately, and lets abilities evolve between seasons. Treating the
components as independent proves untenable, since they compete for a common
budget of possessions. Fitted to a new dataset of 940 matches, the model
identifies the directly relegated club from the ninth round, before the league
table does, predicts the play-out field better than the table, and shows the
play-out itself to be close to a coin toss. Promoted clubs start below the
league on every component except drawing exclusions: knowing a club was
promoted improves forecasts; knowing how does not.

\end{abstract}

\section{Introduction}
Water polo is one of the great traditional team sports. The men's tournament
has been contested at the modern Olympic Games since Paris 1900, making it,
with football, the longest-standing team sport on the Olympic programme. Its history is intertwined with
that of European sport more broadly:
Hungary's nine Olympic titles and its storied rivalries remain a touchstone of
the country's sporting identity, Serbia won gold at the last three Games, and the club game has deep roots throughout Southern Europe and
the Balkans, with Italy, Spain, Croatia, Montenegro and Greece sustaining fully
professional leagues. The sport has an equally long, if distinct, tradition in
the United States, where it is a classic intercollegiate discipline: the NCAA men's
championship has been contested since 1969, and the collegiate game remains the
centre of American water polo.

Despite this pedigree, water polo is almost invisible in the statistical
literature on sport. Quantitative work on the game is active, but it belongs to
the sport-science tradition of notational analysis, which relates game-related
statistics to match outcomes descriptively \citep[e.g.][]{lupo2010,
escalante2011, lupo2025}. Model-based treatments amount to illustrative
examples. \citet{karlis2003} use a water polo dataset among the illustrations
of their bivariate Poisson models, and a short worked example on water polo
goals appears in \citet{ntzoufras2009}. We are not aware of a hierarchical
model tailored to the structure of the sport, or of an analysis of a full
league season.

The contrast with other team sports is sharp. In football, a line of
work running from \citet{maher1982} through \citet{dixon1997} and
\citet{karlis2003} established Poisson models for goal counts, and
\citet{baio2010} popularised the Bayesian hierarchical formulation. Dynamic
extensions track team strength over time \citep{glickman1998, koopman2015,
cattelan2013}, and joint models handle goals alongside disciplinary events
\citep{titman2015}. \citet{ridall2025} embed English league football in a
single state-space model spanning four divisions, giving particular attention
to newly promoted sides. Beyond football, volleyball
\citep{egidi2020, gabrio2021}, tennis \citep{ingram2019, leonelli2026exceptional},
rugby \citep{fioravanti2023} and biathlon \citep{leonelli2026predicting} have all
been treated hierarchically. These papers share the feature we exploit
here. The sport records the components of scoring, and a model that respects
that structure learns more than one fitted to the final score alone.

Water polo suits this treatment particularly well. Its goals arise from three
tactically distinct situations, and official match reports record all three
separately. Even-strength goals, called action goals, are scored six
against six in open play. Exclusion fouls temporarily reduce the defending side
to five field players, and the resulting man-up situations, of which a team
contests on average around a dozen per match, are converted at rates that vary
markedly across teams. Penalty fouls give rise to direct shots from five
metres, awarded roughly twice per team per match and converted around three
quarters of the time. Official match reports record all three components
separately, including the number of man-up opportunities and penalties awarded,
so that opportunity generation and conversion can be modelled as distinct
skills. The decomposition is the water polo analogue of modelling tries and
kicks separately in rugby or of distinguishing even-strength from power-play
scoring in ice hockey, and it turns a match forecast into an interpretable
tactical profile: a struggling team may be undisciplined, poor at defending
the man-down, wasteful with the extra player, or simply weak six against six,
and these diagnoses have different remedies.

Our substantive question concerns the bottom of the table. In contemporary
European water polo the championship race, the object of most
sports forecasting exercises, has ceased to be a question. In Italy,
Pro Recco won the 2025--26 Serie A1 for its fifth consecutive title and its
thirty-eighth overall. In Spain, Atl\`etic-Barceloneta took the 2025--26
Divisi\'on de Honor for its twenty-first consecutive league title. In Hungary,
Ferencv\'aros has won the last five editions of the OB~I, twenty-eight in all,
at one point accumulating one hundred consecutive domestic victories.
Model-based estimates of title probabilities in such leagues
are close to degenerate. The statistically and economically live questions sit
at the bottom of the standings: which clubs will be relegated, and how much should be
believed, in advance, about the newly promoted clubs who are relegation's most
frequent protagonists. Newly promoted sides pose a well-known cold-start
problem for dynamic strength models. The standard remedies initialise
newcomers at a common learned mean \citep{ridall2025}, guard against
over-shrinkage of extreme teams through mixture priors \citep{baio2010}, or
inflate initial variances in rating systems \citep{glickman1999}. Evidence on
whether lower-division performance carries usable information is mixed:
\citet{ridall2025} find that English promoted sides are barely distinguishable
a priori, plausibly because promotion triggers wholesale squad turnover. Water polo offers an informative contrast. Rosters are smaller and transfer
markets thinner, so squad continuity across promotion is plausibly greater, and
a club's second-division record might be a more credible signal of its
first-division strength. We test that expectation, and our results do not
support it.

This paper makes four contributions. First, we assemble and release a new
dataset covering the five Serie A1 seasons from 2021--22 to 2025--26, scraped
from official match reports through a documented protocol, together with a
register of the ten clubs promoted into the league over that window. No
comparable data at this level of detail has been available for the sport.
Second, we build a model that uses that detail: a Bayesian hierarchical
formulation, the first designed for water polo, in which action goals, man-up
opportunities and conversions, and penalties awarded and converted are modelled
jointly, with correlated team abilities, component-specific home advantages and
strengths that evolve across seasons. Posterior predictive checking shows that
treating the components as conditionally independent is untenable, because they
compete for a common budget of possessions, and we add a term that corrects it.
Third, we use the model to study the relegation decision stage by stage, rather
than reporting one measure of predictive accuracy, and report how much can be
known at each stage. Fourth, we compare prior strategies for newly promoted
clubs, covering exchangeable learned priors, lower-division-informed priors,
covariate-based priors and carry-over priors for returning clubs.



\section{Water polo, the league and the data}\label{sec:data}

A water polo match is contested by two teams of seven players, six field
players and a goalkeeper, over four periods of eight minutes of effective play,
with a thirty-second shot clock governing each possession. Physical contact is
policed through a graduated system of fouls, two of which shape the structure
of scoring. An exclusion foul removes the offending player for twenty seconds,
during which the attacking side plays with a numerical advantage; these man-up
situations are the principal special situation of the game, averaging eleven
attempts per team per match across our window, with season means rising from
9.8 to 12.4, and converted thirty-eight per cent of the time. A penalty foul, sanctioning the
denial of a probable goal inside the five-metre line, awards a direct shot on
goal; penalties are far rarer, 1.4 per team per match on average across the window,
and are converted at eighty per cent. All remaining goals are
scored at even strength and are conventionally labelled action goals.

The official match report records, for each side, the final and
per-period scores, every goal with its type, the number of man-up situations
with the number converted, and the number of penalties awarded with their
outcomes. The decomposition
$\text{goals} \;=\; \text{action goals} \;+\; \text{man-up goals} \;+\;
\text{penalty goals}$ 
is therefore observable match by match, together with the opportunity counts
that underlie the two special-situation components. It is this richness,
routinely available yet never exploited, that the model of
Section~\ref{sec:model} is built around.

The Serie A1 is the top tier of Italian men's water polo, contested by fourteen
clubs, with promotion to and relegation from the Serie A2 below it. Our window
covers the five seasons from 2021--22 to 2025--26. Three of them followed a
home-and-away round robin of twenty-six rounds; two, 2021--22 and 2023--24,
split the league after a thirteen-round first leg into a championship pool and
a relegation pool of seven clubs each, the first by emergency after a wave of
Covid-19 cases and the second by design. The title is decided by play-off
series among the leading clubs. Relegation combines a direct place for the
club finishing last with a play-out bracket among the four clubs above it, in
which the losers of the two semifinals meet and the loser of that final goes
down as well. Points were three for a win and one for a draw until 2024--25;
from 2025--26 no league match may end level, and a penalty shoot-out follows
every regulation draw. Appendix~\ref{app:formats} sets out the season-by-season
detail, which matters because the format changes the mapping from team strength
to relegation risk. Our model is specified at the level of match scoring, so
any format can be layered on top by simulation, as
Section~\ref{sec:promoted-eval} requires.

Table~\ref{tab:window} summarises the window. Twenty-one
clubs appear, eight of them ever-present, and two enter by promotion every
year. The pattern at the top could hardly be starker: Pro Recco won all five
titles and every regular season except 2024--25, when it finished level with
Brescia and lost first place on head-to-head record. The bottom changed hands
every year. Sixteen of the twenty-one clubs occupied a direct relegation place
or contested a play-out at least once, and in three of the last four seasons a
newly promoted club went straight back down.

\begin{table}[t]
\centering
\small
\caption{The league over the study window. Left: the twenty-one clubs, a
bullet marking participation and an arrow marking entry by promotion from the
Serie A2, all ten promotion episodes having come through the Serie A2
play-offs. Right: the classification of each season, with the play-off,
play-out and relegation outcomes as published. Club codes are shared between
the two panels.}
\label{tab:window}
\begin{minipage}[t]{0.44\textwidth}\centering
\resizebox{\linewidth}{!}{
\begin{tabular}{llccccc}
\toprule
Club & Code & 2021--22 & 2022--23 & 2023--24 & 2024--25 & 2025--26 \\
\midrule
Brescia & BRE & $\bullet$ & $\bullet$ & $\bullet$ & $\bullet$ & $\bullet$ \\
Ortigia & ORT & $\bullet$ & $\bullet$ & $\bullet$ & $\bullet$ & $\bullet$ \\
Posillipo & POS & $\bullet$ & $\bullet$ & $\bullet$ & $\bullet$ & $\bullet$ \\
Pro Recco & REC & $\bullet$ & $\bullet$ & $\bullet$ & $\bullet$ & $\bullet$ \\
Quinto & QUI & $\bullet$ & $\bullet$ & $\bullet$ & $\bullet$ & $\bullet$ \\
Savona & SAV & $\bullet$ & $\bullet$ & $\bullet$ & $\bullet$ & $\bullet$ \\
Telimar & TEL & $\bullet$ & $\bullet$ & $\bullet$ & $\bullet$ & $\bullet$ \\
Trieste & TRI & $\bullet$ & $\bullet$ & $\bullet$ & $\bullet$ & $\bullet$ \\
\addlinespace
Catania & CAT & $\uparrow$ & $\bullet$ & $\bullet$ & $\bullet$ & \\
De Akker & DEA & & $\uparrow$ & $\bullet$ & $\bullet$ & $\bullet$ \\
Onda Forte Roma & OFR & $\bullet$ & $\bullet$ & $\bullet$ & $\bullet$ & \\
Salerno & SAL & $\bullet$ & $\bullet$ & $\bullet$ & & $\uparrow$ \\
\addlinespace
Roma Vis Nova & RVN & & & $\uparrow$ & $\bullet$ & $\bullet$ \\
\addlinespace
Anzio & ANZ & $\uparrow$ & $\bullet$ & & & \\
Florentia & FLO & & & & $\uparrow$ & $\bullet$ \\
Olympic Roma & OLY & & & & $\uparrow$ & $\bullet$ \\
\addlinespace
Bogliasco & BOG & & $\uparrow$ & & & \\
Camogli & CAM & & & $\uparrow$ & & \\
CC Napoli & CCN & & & & & $\uparrow$ \\
Lazio & LAZ & $\bullet$ & & & & \\
Metanopoli & MET & $\bullet$ & & & & \\
\bottomrule
\end{tabular}}
\end{minipage}\hfill
\begin{minipage}[t]{0.54\textwidth}\centering
\resizebox{\linewidth}{!}{
\begin{tabular}{r*{5}{lr}}
\toprule
 & \multicolumn{2}{c}{2021--22} & \multicolumn{2}{c}{2022--23} &
   \multicolumn{2}{c}{2023--24} & \multicolumn{2}{c}{2024--25} &
   \multicolumn{2}{c}{2025--26} \\
\midrule
1 & REC & 39 & REC & 76 & REC & 39 & BRE & 73 & REC & 78 \\
2 & BRE & 33 & BRE & 73 & SAV & 36 & REC & 73 & BRE & 72 \\
3 & TRI & 29 & ORT & 59 & BRE & 30 & SAV & 66 & SAV & 65 \\
4 & TEL & 28 & TEL & 56 & ORT & 27 & TRI & 49 & POS & 45 \\
5 & ORT & 28 & TRI & 49 & TEL & 25 & DEA & 46 & TRI & 44 \\
6 & SAV & 27 & SAV & 44 & TRI & 21 & POS & 38 & QUI & 38 \\
7 & SAL & 18 & QUI & 37 & DEA & 17 & RVN & 37 & DEA & 35 \\
8 & QUI & 18 & POS & 27 & QUI & 16 & ORT & 35 & RVN & 33 \\
9 & MET & 12 & OFR & 27 & POS & 16 & QUI & 29 & OLY & 29 \\
10 & ANZ & 11 & ANZ & 26 & OFR & 14 & TEL & 28 & CCN & 28 \\
11 & POS & 10 & DEA & 21 & CAT & 10 & FLO & 26 & ORT & 24 \\
12 & CAT & 7 & SAL & 19 & RVN & 8 & OLY & 18 & TEL & 23 \\
13 & OFR & 7 & CAT & 13 & SAL & 7 & CAT & 8 & SAL & 18 \\
14 & LAZ & 1 & BOG & 5 & CAM & 0 & OFR & 5 & FLO & 14 \\
\midrule
Champions & \multicolumn{2}{c}{REC} & \multicolumn{2}{c}{REC} & \multicolumn{2}{c}{REC} & \multicolumn{2}{c}{REC} & \multicolumn{2}{c}{REC} \\
Runners-up & \multicolumn{2}{c}{BRE} & \multicolumn{2}{c}{BRE} & \multicolumn{2}{c}{SAV} & \multicolumn{2}{c}{BRE} & \multicolumn{2}{c}{BRE} \\
Third & \multicolumn{2}{c}{SAV} & \multicolumn{2}{c}{ORT} & \multicolumn{2}{c}{BRE} & \multicolumn{2}{c}{SAV} & \multicolumn{2}{c}{--} \\
Relegated & \multicolumn{2}{c}{LAZ, MET} & \multicolumn{2}{c}{BOG, ANZ} & \multicolumn{2}{c}{CAM, SAL} & \multicolumn{2}{c}{OFR, CAT} & \multicolumn{2}{c}{FLO, SAL} \\
\bottomrule
\end{tabular}}
\end{minipage}
\end{table}

We scraped the data from the official match centre of the Italian Swimming
Federation, whose pages record every goal by type together with each side's
man-up attempts and penalties awarded, and validated every record against the
published classifications. Nothing comparable has previously been available
for this sport. Appendix~\ref{app:data} describes the sources, the extraction
protocol and the checks.


\section{A multi-component dynamic hierarchical model}\label{sec:model}

\subsection{Observation model}\label{sec:obs}

Index team-match records by $i = 1, \dots, N$, with two records per match, and
write $t(i)$ for the team, $o(i)$ for the opponent, $s(i)$ for the season and
$h_i \in \{0, 1\}$ for the home indicator. For record $i$ we observe the
action goals $Y_i$, the man-up attempts $M_i$ and conversions $G_i$, and the
penalties awarded $R_i$ and scored $Z_i$, all counted in regulation time; a
shoot-out, where one occurs, is outside the model. The observation model treats
opportunities as counts and conversions as conditionally binomial:
\begin{align}
Y_i &\sim \Poisson\{\lambda^{(a)}_i\}, &
M_i &\sim \Poisson\{\lambda^{(x)}_i\}, &
G_i \mid M_i &\sim \Binomial\{M_i,\, p^{(x)}_i\}, \label{eq:obs1}\\
R_i &\sim \Poisson\{\lambda^{(p)}_i\}, &
Z_i \mid R_i &\sim \Binomial\{R_i,\, p^{(p)}_i\}, \label{eq:obs2}
\end{align}
with the five records conditionally independent given the linear predictors
\begin{align}
\log \lambda^{(a)}_i &= \mu^{(a)}_{s(i)} + \eta^{(a)} h_i +
  \alpha^{(a)}_{t(i), s(i)} + \beta^{(a)}_{o(i), s(i)}
  + \boldsymbol{\gamma}^\top \tilde{\boldsymbol{x}}_i, \label{eq:lp-a}\\
\log \lambda^{(x)}_i &= \mu^{(x)}_{s(i)} + \eta^{(x)} h_i +
  \alpha^{(x)}_{t(i), s(i)} + \beta^{(x)}_{o(i), s(i)}, \label{eq:lp-x}\\
\log \lambda^{(p)}_i &= \mu^{(p)}_{s(i)} + \eta^{(p)} h_i +
  \alpha^{(p)}_{t(i), s(i)} + \beta^{(p)}_{o(i), s(i)}, \label{eq:lp-p}\\
\logit\, p^{(x)}_i &= \mu^{(c)}_{s(i)} + \eta^{(c)} h_i +
  c_{t(i), s(i)} + k_{o(i), s(i)}, \label{eq:lp-c}\\
\logit\, p^{(p)}_i &= \mu^{(w)}_{s(i)} + \eta^{(w)} h_i
  + w_{t(i), s(i)}. \label{eq:lp-w}
\end{align}
Here $\mu^{(\cdot)}_{s}$ are season-specific league-level intercepts and
$\eta^{(\cdot)}$ component-specific home advantages. The intercepts are
indexed by season because the scoring level of the league is not stable over
the window, as Section~\ref{sec:results-eda} documents; holding them fixed
would force team abilities to absorb a league-wide drift.
The term $\boldsymbol{\gamma}^\top \tilde{\boldsymbol{x}}_i$ in
\eqref{eq:lp-a}, where $\tilde{\boldsymbol{x}}_i$ standardises the team's own
man-up attempts $M_i$ and penalties awarded $R_i$ in that match, is discussed
in Section~\ref{sec:substitution}. Each team-season carries an attacking
ability vector $\boldsymbol{\alpha}_{ts} = (\alpha^{(a)}_{ts},
\alpha^{(x)}_{ts}, \alpha^{(p)}_{ts})^\top$, collecting even-strength scoring,
the propensity to draw exclusions and the propensity to draw penalties, and a
defensive vector $\boldsymbol{\beta}_{ts}$ with the corresponding conceding
propensities, positive values indicating weaker defence. The scalar effects
$c_{ts}$, $k_{ts}$ and $w_{ts}$ capture man-up shooting, man-down defending
and penalty shooting. The decomposition confers three advantages over
modelling the total score: conversion skill is separated from opportunity
generation, using the observed denominators rather than latent ones;
information is pooled coherently, since a team's dozen man-up attempts per
match inform its conversion parameter far more strongly than its two penalties
inform penalty shooting; and every forecast is accompanied by an interpretable
tactical profile. Total goals are recovered as $Y_i + G_i + Z_i$, and any
result-level quantity, including league points under a given format, follows
by simulation.

Within a season abilities are held constant, a simplification supported at
this granularity by the round-by-round evaluation of
Section~\ref{sec:results-predictions}, in which refitting after every third
round does not overturn the ordering of clubs. We considered within-season dynamics in the manner of \citet{koopman2015} or
\citet{ridall2025} and did not adopt them here. With twenty-six rounds and
fourteen clubs, a state that evolves every round is weakly identified, and the
round-by-round tier already lets the posterior update as a season proceeds at a
fraction of the computational cost. Section~\ref{sec:discussion} returns to
the point.

\subsection{Component substitution}\label{sec:substitution}

Treating the components as conditionally independent is the natural starting
point, but it has a cost. A possession that ends in a drawn exclusion is a
possession that did not end in an even-strength shot. The components compete
for a common budget of possessions, so within a match more man-up
opportunities should come partly at the expense of action opportunities. We
call this substitution, and we capture it with the term
$\boldsymbol{\gamma}^\top \tilde{\boldsymbol{x}}_i$ in \eqref{eq:lp-a}, which
lets the even-strength rate fall as the team's own opportunity counts rise. A model that correlates
abilities across components, as formalised in \eqref{eq:lkj} below, can only
add positive dependence between them. It cannot produce the negative dependence that
substitution implies. Section~\ref{sec:results-model} shows that this matters.
Without the term, the model reproduces about three times the covariance between
components that we observe, and so over-disperses the match
total, even though each component on its own is well calibrated.

Because the opportunity counts are themselves modelled, conditioning the
action rate on them is a factorisation of the joint distribution rather than a
double use of the data: the model specifies $P(M_i, R_i)$, then $P(Y_i \mid
M_i, R_i)$, then the conversions given the opportunities, and simulation
proceeds in the same order, so the model remains fully generative and can
still forecast an unplayed match. We take $\tilde{\boldsymbol{x}}_i =
\{(M_i - \bar{M})/s_M,\, (R_i - \bar{R})/s_R\}^\top$ with $\boldsymbol{\gamma}
\sim \Normal_2(\boldsymbol{0}, 0.5^2 I)$, expecting both coefficients to be
negative.

A copula, or a multivariate count distribution such as the bivariate Poisson of
\citet{karlis2003}, would also induce dependence between the components. We
prefer the conditional form for three reasons. It has a mechanism behind it, so
the sign of $\boldsymbol{\gamma}$ is a testable claim about how possessions end
rather than a free association parameter. It conditions on counts the model
already generates, so simulation remains a simple forward pass and the model
can still forecast an unplayed match. The common alternatives also carry the
limitation we need to avoid: the bivariate Poisson admits only non-negative
correlation, and a copula on the three margins would fit the dependence
without explaining it. Section~\ref{sec:results-predictions} reports that the
term also improves out-of-sample scores within a season, so it is not merely a
diagnostic repair.

\subsection{Cross-sectional structure}\label{sec:crosssec}

Within a season, the attacking vectors are modelled as exchangeable across
teams with a correlated prior,
\begin{equation}\label{eq:lkj}
\boldsymbol{\alpha}_{ts} \sim \Normal_3\{\boldsymbol{m}^{\alpha}_{ts},\,
\Sigma_\alpha\}, \qquad
\Sigma_\alpha = D_\alpha \Omega_\alpha D_\alpha,
\end{equation}
with $D_\alpha$ the diagonal matrix of component scales, $\Omega_\alpha$ a
correlation matrix with an LKJ prior \citep{lkj2009}, and the mean
$\boldsymbol{m}^{\alpha}_{ts}$ determined by the dynamics of
Section~\ref{sec:dynamics} for continuing clubs and by the strategies of
Section~\ref{sec:promoted} for newly promoted ones. The defensive vectors
$\boldsymbol{\beta}_{ts}$ carry the analogous structure with scales $D_\beta$
and correlation $\Omega_\beta$, and the conversion effects are exchangeable
normal with their own scales. The correlations encode the tactical hypothesis
that offensive pressure manifests jointly, as even-strength scoring, drawn
exclusions and drawn penalties, and its defensive mirror; the extent to which
five seasons of data can identify them is itself of interest. League intercepts
absorb overall levels, so the team effects are softly identified through their
zero-centred population distributions, with sum-to-zero summaries reported for
interpretation.

\subsection{Dynamics across seasons}\label{sec:dynamics}

For a club appearing in consecutive seasons, abilities evolve by a stationary
first-order autoregression towards the league mean,
\begin{equation}\label{eq:ar1}
\boldsymbol{\alpha}_{t, s+1} \mid \boldsymbol{\alpha}_{t, s} \sim
\Normal_3\{P_\alpha \boldsymbol{\alpha}_{t, s},\,
(I - P_\alpha^2)^{1/2}\, \Sigma_\alpha\, (I - P_\alpha^2)^{1/2}\},
\end{equation}
with $P_\alpha = \mathrm{diag}(\rho^{(a)}_\alpha, \rho^{(x)}_\alpha,
\rho^{(p)}_\alpha)$ and persistence parameters $\rho \in (0, 1)$, so that the
implied marginal distribution of abilities is the cross-sectional prior
\eqref{eq:lkj} in every season and the model nests both full persistence and
season-by-season independence. Defensive vectors and the scalar conversion
effects evolve analogously, with component-specific persistence, reflecting
the possibility that special-situation skills, which are heavily
system-dependent, regress faster under coaching turnover than does raw
even-strength quality. The estimates of Section~\ref{sec:results-model}
justify the separate treatment: the conversion effects are as persistent as
the action block, near $0.83$, while the penalty opportunity dimensions are
markedly less so, near $0.50$, so a shared persistence parameter would be
misspecified in both directions.

\subsection{Prior distributions and computation}\label{sec:priors}

League intercepts receive normal priors anchored at the pooled observed rates,
six action goals, eleven man-up attempts and one and a half penalties per
team-match, with conversion intercepts anchored at thirty-eight and eighty-one
per cent on the logit scale. Home advantages and the substitution coefficients
receive zero-centred normal priors, scales receive half-normal priors,
correlation matrices receive LKJ priors with shape two, and persistence
parameters receive normal priors truncated to the stationary region and
favouring moderate to high persistence. Appendix~\ref{app:model} gives the
full specification. Posterior computation uses Hamiltonian Monte Carlo as implemented in Stan
\citep{stan2017} with a non-centred parametrisation of all hierarchical
layers. Four chains of three thousand iterations, half discarded as warm-up,
were run for every specification we report. Convergence was satisfactory
throughout: the largest potential scale reduction factor was $1.007$, the
smallest bulk effective sample size $1047$, and the one specification that produced
divergent transitions was refitted at a higher target acceptance rate. Model assessment combines approximate leave-one-out cross-validation
\citep{vehtari2017} for structural comparisons with the strictly out-of-sample
forward-chaining design of Section~\ref{sec:promoted-eval} for the questions
involving promoted clubs.

\subsection{The model set and its evaluation}\label{sec:modelset}

The specification above involves several choices that we test rather than
assume: whether the decomposition improves on a model of match totals, whether
abilities need to evolve, whether the component blocks need to be correlated,
whether every component supports a team-level effect, and whether the
substitution term is required. We therefore estimate a set of configurations that isolate these choices one at a time and
compare them out of sample. Table~\ref{tab:models} lists the eleven configurations
carried through to evaluation: a double Poisson on match totals in the manner
of \citet{maher1982} with static, autoregressive and correlated variants; the
decomposed model of \eqref{eq:obs1} and \eqref{eq:obs2}; variants removing the
penalty shooting effect, forcing a single home advantage and adding the
substitution term; and the promoted-team priors of
Section~\ref{sec:promoted}.

\begin{table}[t]\centering\small
\caption{The eleven configurations carried to out-of-sample evaluation. All
dynamic configurations use stationary first-order autoregressive team effects.
\textsc{sub} is the specification used for all substantive results.}
\label{tab:models}
\begin{tabular}{lp{7.2cm}ccc}\toprule
label & specification & decomposed & substitution & entrant prior \\\midrule
\textsc{tot-s} & match totals, static abilities & -- & -- & -- \\
\textsc{tot-d} & match totals, dynamic abilities & -- & -- & -- \\
\textsc{tot-dc} & match totals, dynamic and correlated abilities & -- & -- & -- \\
\textsc{dec} & decomposed, dynamic and correlated abilities & yes & -- & -- \\
\textsc{dec-h} & \textsc{dec} with a single home advantage & yes & -- & -- \\
\textsc{dec-p} & \textsc{dec} without a penalty shooting effect & yes & -- & -- \\
\textsc{dec-sub} & \textsc{dec} with the substitution term & yes & yes & -- \\
\textsc{sub} & \textsc{dec-p} with the substitution term & yes & yes & -- \\
\textsc{sub-s1} & \textsc{sub} with entrant prior & yes & yes & S1 \\
\textsc{sub-s2} & \textsc{sub} with entrant prior & yes & yes & S2 \\
\textsc{sub-s3} & \textsc{sub} with entrant prior & yes & yes & S3 \\
\bottomrule\end{tabular}\end{table}

Comparison proceeds on two tiers, which answer different questions and, as
Section~\ref{sec:results-predictions} shows, give different answers. Tier one
is season-ahead forward chaining: the model is fitted to seasons
$1, \dots, k$ and the whole of season $k+1$ is predicted, for $k = 1, \dots,
4$. Here the coming season's intercepts are unknown and are carried forward
from the last observed season, which is the position of a forecaster before a season begins. Tier two is round-by-round: within the final season
the model is refitted after every third round and the next block predicted,
so that the league's scoring level and the clubs' current states are both
informed by data. Scores are computed on a common scale for every
configuration, namely the total goals of a team-match and the ordered match
result, so that totals and decomposed models remain comparable even though their
likelihoods are not. Between any two configurations whose likelihoods cover the
same observations, approximate leave-one-out
cross-validation \citep{vehtari2017} supplements the out-of-sample scores.

\section{Prior strategies for newly promoted teams}\label{sec:promoted}

\subsection{The problem}\label{sec:promoted-problem}

Equation~\eqref{eq:ar1} propagates information for continuing clubs, but a
club entering the league has no state to propagate. In our window, two clubs
enter each season, and, because entrants are drawn into the relegation battle far more often
than not, the initialisation choice bears directly on the quantities of
substantive interest. The problem has received scattered treatment: rating
systems inflate the initial variance of new entrants \citep{glickman1999};
hierarchical football models guard against over-shrinkage of teams far from
the league mean through mixture priors \citep{baio2010}; and, most directly,
\citet{ridall2025} initialise promoted English clubs at a common estimated
mean, finding lower-division form to be barely informative and attributing
this to wholesale squad reconstruction upon promotion. We formalise a family
of strategies that contains these proposals as special cases and extends them
in directions that the structure of water polo makes both feasible and
interesting.

Let $t$ be a club entering the league in season $s_0$, write
$\boldsymbol{\theta}_{t s_0} = (\boldsymbol{\alpha}_{t s_0}^\top,
\boldsymbol{\beta}_{t s_0}^\top, c_{t s_0}, k_{t s_0}, w_{t s_0})^\top$ for
its full ability state, and let $\mathcal{P}$ denote the set of promotion
episodes in the data: all ten entries of the window, including the two clubs
promoted into its opening season. In season one, entrants draw the entrant
prior below while incumbent clubs draw the stationary prior, mirroring the
information the league itself held at that point and lending $\mathcal{P}$
its full size where the entrant-prior parameters are weakly identified. All
strategies specify
$\boldsymbol{\theta}_{t s_0} \sim \Normal\{\boldsymbol{m}_t,\, \Sigma_P\}$
and differ in the construction of $\boldsymbol{m}_t$ and in what is learned
across $\mathcal{P}$.

\subsection{The strategies}\label{sec:promoted-strategies}

\paragraph{S1: learned exchangeable prior.} All entrants share a common mean,
$\boldsymbol{m}_t = \boldsymbol{m}_P$, with $\boldsymbol{m}_P$ and $\Sigma_P$
given hyperpriors and estimated jointly with the rest of the model. This is
the multivariate, fully Bayesian analogue of the initialisation of
\citet{ridall2025}: the data determine both how far below the league mean an
average entrant sits, component by component, and how variable entrants are.
The league-average prior with inflated variance is the special case
$\boldsymbol{m}_P = \boldsymbol{0}$.

\paragraph{S2: lower-division-informed prior.} The entrant's mean is regressed
on its Serie A2 record, $\boldsymbol{m}_t = \boldsymbol{m}_P + \Lambda
\boldsymbol{z}_t$, where $\boldsymbol{z}_t$ collects standardised summaries of
the promotion season and $\Lambda$ is estimated across $\mathcal{P}$. The
vector $\boldsymbol{z}_t$ contains the A2 points share and goal difference per
game, the finishing position, and an indicator of whether the club came from
the northern or the southern group. The matrix $\Lambda$ then measures,
component by component, how much second-division form transfers, providing a
direct water polo counterpart to the negative English finding of
\citet{ridall2025}. The assembled register
shows why the transfer cannot be taken at face value.
Entrants converted between $34.6$ and $48.8$ per cent of their second-division
man-up situations, pooling at $41.8$ per cent against a Serie A1 norm of
$37.9$, and between $85.7$ and $97.9$ per cent of their penalties, pooling at
$91.1$ against an A1 norm of $80.2$ (Table~\ref{tab:register}). Second-division rates therefore arrive against weaker
defences, and $\Lambda$ must absorb a divisional level shift, larger for
penalties than for man-up play, as well as the ordering of clubs.

\paragraph{S3: carry-over for returning clubs.} Some entrants are returning after
relegation rather than arriving for the first time. In our window this applies
to one club: Salerno played the first three seasons, was relegated after
2023--24 and returned for 2025--26, so $\Delta_t = 1$. Every other entrant has no Serie A1 record inside the window, and for
them the carry-over degenerates to S1. The strategy is therefore identified
from a single episode, a limitation we return to in
Section~\ref{sec:discussion}. For a
club with a previous Serie A1 state $\boldsymbol{\theta}_{t, s_e}$ recorded
$\Delta_t$ seasons earlier, the evolution \eqref{eq:ar1} is applied over the
gap with an additional inflation for the unobserved spell,
\begin{equation}\label{eq:carryover}
\boldsymbol{m}_t = P^{\Delta_t} \boldsymbol{\theta}_{t, s_e}, \qquad
\Sigma_P(\Delta_t) = (I - P^{2 \Delta_t})^{1/2} \Sigma (I -
P^{2 \Delta_t})^{1/2} + \Sigma_{\text{gap}},
\end{equation}
so that a short absence retains substantial information while a long absence
collapses towards S1, into which \eqref{eq:carryover} degenerates as
$\Delta_t$ grows. The strategy is combinable with S2, the two information
sources entering through precision weighting.

\subsection{Design of the comparison}\label{sec:promoted-eval}

The strategies are compared under the evaluation design of
Section~\ref{sec:modelset}. Two features are specific to this contest. The
primary comparison restricts attention to matches involving entrants, and
within them to the opening rounds, where the strategies differ most and where
the entrant prior has not yet been overwritten by the club's own results. The
fitted quantities are also of interest in their own right: $\Lambda$ measures
how far second-division form transfers, and the comparison of entrant
posteriors under S1 against their realised trajectories shows how quickly the
league teaches the model what a newcomer is.


\section{Results}\label{sec:results}

\subsection{Exploratory analysis}\label{sec:results-eda}

Table~\ref{tab:descriptives} describes the assembled data.
The most striking feature of the window is a pronounced rise in scoring: total
goals per team-match move from $9.91$ in 2021--22 to $13.49$ in 2025--26, an
increase of thirty-six per cent in five seasons. The decomposition shows at
once that this is not a change in finishing. Man-up conversion is flat across
the window, between $37.5$ and $38.9$ per cent, and penalty conversion varies
without trend between $75.9$ and $83.8$ per cent. What changes is the supply
of opportunities. Man-up attempts rise from $9.79$ to $12.38$ per team-match
and penalties awarded more than double, from $0.96$ to $2.00$. Action goals
themselves rise from $5.54$ to $7.29$, but their share of the total falls
from $55.8$ to $53.9$ per cent while the penalty share rises from $7.8$ to
$11.3$. The discontinuity has a cause. New playing rules took effect in 2025--26: a
shorter field, a shorter possession clock and a wider definition of the penalty
foul. More possessions per match and more penalties are exactly what those
changes were designed to produce.

\begin{table}[t]
\centering
\caption{Per team-match means by season over played matches, the 2022--23
forfeit excluded. Man-up and penalty entries report conversions over
opportunities; the final row counts matches decided at a shoot-out.}
\label{tab:descriptives}
\small
\begin{tabular}{lccccc}
\toprule
 & 2021--22 & 2022--23 & 2023--24 & 2024--25 & 2025--26 \\
\midrule
Matches & 155 & 207 & 158 & 210 & 209 \\
Goals & 9.91 & 10.03 & 10.46 & 11.07 & 13.49 \\
Action goals & 5.54 & 5.33 & 5.16 & 5.67 & 7.29 \\
Man-up (conv/att) & 3.62/9.79 & 3.82/10.08 & 4.24/11.32 & 4.15/10.72 & 4.68/12.38 \\
Man-up conversion (\%) & 37.5 & 37.9 & 37.7 & 38.9 & 38.0 \\
Penalties (scored/awarded) & 0.77/0.96 & 0.89/1.09 & 1.05/1.28 & 1.24/1.48 & 1.52/2.00 \\
Penalty conversion (\%) & 79.7 & 82.0 & 82.2 & 83.8 & 75.9 \\
Shoot-outs & 2 & 3 & 3 & 2 & 16 \\
\bottomrule
\end{tabular}
\end{table}

Teams are not shooting better. They are getting more chances, and more of the
most valuable kind. A model of match totals would record the rise and assign it
to the teams. Our decomposition locates it in the opportunity components, and
the season-indexed intercepts of \eqref{eq:lp-a} to \eqref{eq:lp-w} keep it out of the ability estimates.

Home advantage appears in the raw data in every season, from $10.14$ against
$9.68$ goals per team-match in 2021--22 to $13.81$ against $13.15$ in 2025--26.
It is small next to the seasonal drift, and, as
Section~\ref{sec:results-model} shows, it is spread unevenly across the
components.

Table~\ref{tab:register} summarises the promotion register. Nine of the ten entrants won their group or finished
second; only Olympic Roma, promoted from third in 2023--24 with a goal
difference of $+26$, arrived from outside the top two. The divisional level
shift anticipated in Section~\ref{sec:promoted-strategies} is substantial and
larger for penalties than for man-up play: entrants converted $91.1$ per cent
of their second-division penalties against a Serie A1 norm of $80.2$, and
$41.8$ per cent of their man-up situations against $37.9$.

\begin{table}[t]\centering\small
\caption{The promotion register: the ten clubs entering Serie A1 over the study
window, with their second-division record in the season before entry.
Conversion rates are computed from the entrant's own second-division matches.}
\label{tab:register}
\begin{tabular}{llllrrrrrr}
\toprule
club & enters & A2 season & group & pos. & points & GF & GA &
man-up \% & pen. \% \\ \midrule
Anzio & 2021-22 & 2020-21 & Centro & 1 & 28 & 117 & 75 & 35.4 & 87.5 \\
Catania & 2021-22 & 2020-21 & Sud & 1 & 28 & 141 & 85 & 44.7 & 85.7 \\
De Akker & 2022-23 & 2021-22 & Nord & 1 & 61 & 268 & 129 & 40.2 & 90.5 \\
Bogliasco & 2022-23 & 2021-22 & Nord & 2 & 58 & 280 & 162 & 40.8 & 83.9 \\
Roma Vis Nova & 2023-24 & 2022-23 & Sud & 1 & 63 & 326 & 165 & 39.8 & 93.5 \\
Camogli & 2023-24 & 2022-23 & Nord & 1 & 63 & 303 & 167 & 45.7 & 97.9 \\
Florentia & 2024-25 & 2023-24 & Nord & 1 & 62 & 289 & 167 & 45.3 & 86.4 \\
Olympic Roma & 2024-25 & 2023-24 & Sud & 3 & 40 & 212 & 186 & 34.6 & 89.2 \\
CC Napoli & 2025-26 & 2024-25 & Sud & 1 & 62 & 303 & 183 & 48.8 & 87.1 \\
Salerno & 2025-26 & 2024-25 & Sud & 2 & 58 & 326 & 174 & 45.5 & 94.7 \\
\addlinespace
\multicolumn{8}{l}{pooled, second division} & 41.8 & 91.1 \\
\multicolumn{8}{l}{pooled, Serie A1} & 37.9 & 80.2 \\
\bottomrule\end{tabular}\end{table}

\subsection{Model results}\label{sec:results-model}

\begin{figure}[t]\centering
\includegraphics[width=\textwidth]{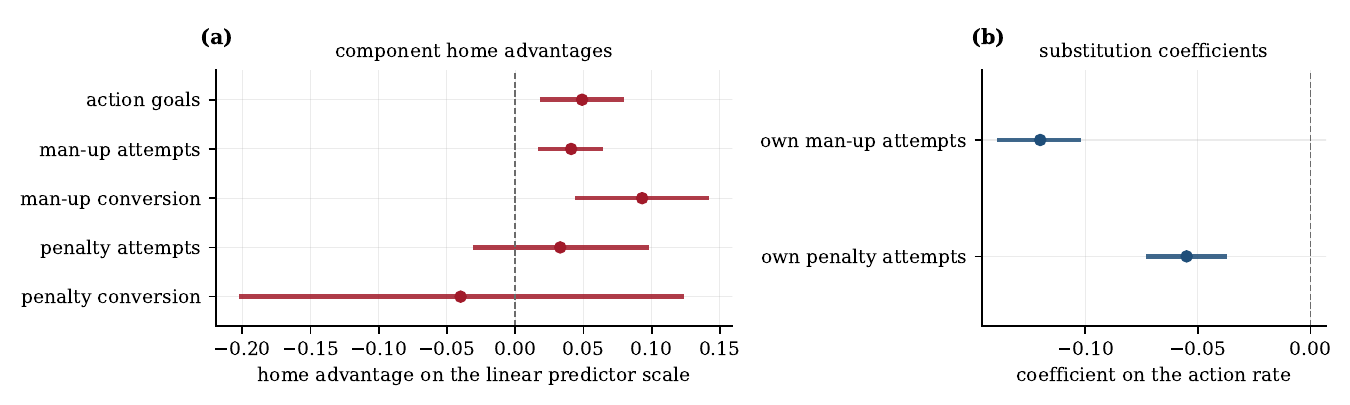}
\caption{Panel (a): the five component home advantages $\eta^{(\cdot)}$ of
\eqref{eq:lp-a} to \eqref{eq:lp-w}, the shift a home side receives on each
component's linear predictor. Panel (b): the two substitution coefficients
$\boldsymbol{\gamma}$ of \eqref{eq:lp-a}, the change in the log action rate
per standard deviation of the side's own man-up attempts and penalties
awarded. Both under \textsc{sub}; posterior means with ninety per cent
credible intervals.}
\label{fig:home}
\alttext{Two panels of posterior means with ninety per cent intervals. Panel (a) shows the home advantage for the five components: largest for man-up conversion, about half that size for action goals and man-up attempts, and indistinguishable from zero for penalties awarded and penalty conversion. Panel (b) shows the two substitution coefficients, both clearly negative.}
\end{figure}

\paragraph{Home advantage.}
The model carries a separate home advantage for each component: the
coefficients $\eta^{(a)}$, $\eta^{(x)}$, $\eta^{(c)}$, $\eta^{(p)}$ and
$\eta^{(w)}$ of \eqref{eq:lp-a} to \eqref{eq:lp-w}, which shift the home
side's action rate, man-up attempts, man-up conversion, penalties awarded and
penalty conversion respectively. Figure~\ref{fig:home}(a) reports their
posteriors under \textsc{sub}. The largest is on man-up conversion,
$\eta^{(c)} = 0.093$ with a ninety per cent interval of $[0.044, 0.142]$ on
the logit scale, an increase in the odds of converting the extra man of about
ten per cent. The action rate carries $\eta^{(a)} = 0.049$ $[0.018, 0.080]$
and the drawing of exclusions $\eta^{(x)} = 0.041$ $[0.017, 0.064]$, both
clear of zero but half the size. Penalties awarded show no home excess,
$\eta^{(p)} = 0.033$ $[-0.031, 0.098]$, and penalty conversion none at all,
$\eta^{(w)} = -0.040$ $[-0.202, 0.124]$. The home effect is largest where the game is most structured and depends most
on set-piece execution. It disappears at the two moments that are most
refereed and most isolated. Configuration \textsc{dec-h} forces one home advantage across
all components. It estimates that advantage at $0.045$, leaves the other four
at their priors, and so loses the finding altogether.

\paragraph{Substitution.}
Both substitution coefficients are clearly negative and precisely estimated:
$-0.120$ $[-0.139, -0.102]$ for the team's own man-up attempts and $-0.055$
$[-0.073, -0.037]$ for its own penalties (Figure~\ref{fig:home}(b)). A
one-standard-deviation increase in exclusions drawn, about two attempts,
lowers the expected action goals by twelve per cent. The mechanism is
substantial rather than marginal, and it is not visible at the level of
totals.

\begin{figure}[t]\centering
\includegraphics[width=\textwidth]{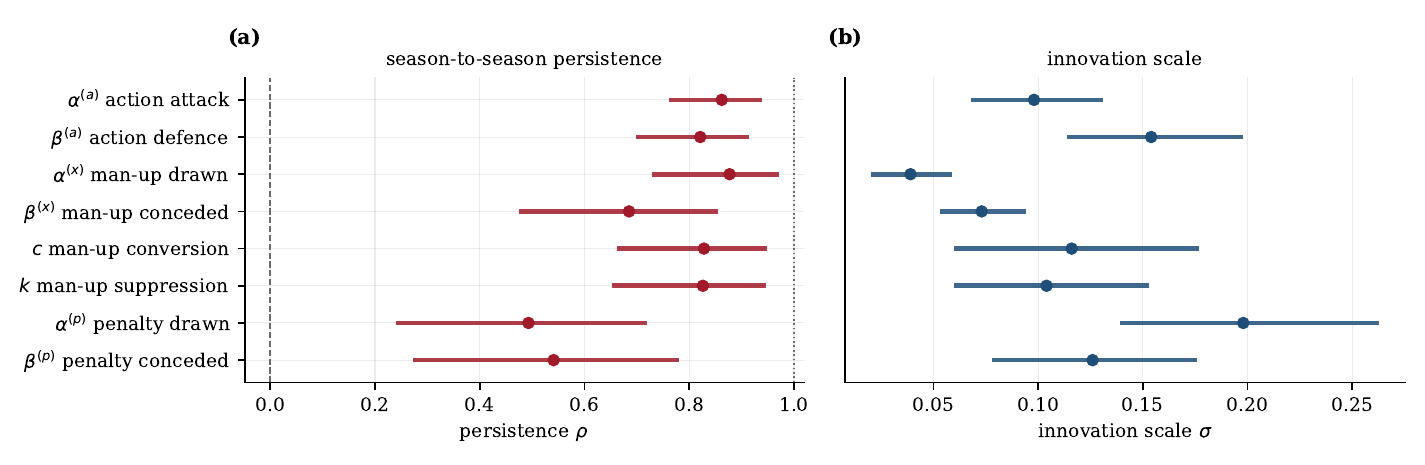}
\caption{Season-to-season persistence and innovation scale by ability
dimension under \textsc{sub}.}
\label{fig:dyn}
\alttext{Two panels of posterior means with ninety per cent intervals across the eight ability dimensions. Panel (a) shows season-to-season persistence, high and tightly estimated for the action and man-up dimensions and around one half with wide intervals for the two penalty dimensions. Panel (b) shows the innovation scales, smallest for man-up attempts drawn and largest for the penalty dimensions.}
\end{figure}

\paragraph{Persistence.}
Figure~\ref{fig:dyn} reports the autoregressive structure. Even-strength ability and the propensity to draw exclusions are
highly persistent, with $\rho$ between $0.82$ and $0.88$, and their
innovations are small. Man-up shooting and man-up suppression are almost as
persistent, near $0.83$. The penalty dimensions are markedly less so:
drawing penalties has $\rho = 0.49$ $[0.24, 0.72]$ and conceding them $0.54$
$[0.27, 0.78]$, with innovation scales two to three times those of the action
block. Penalties are, in this league, closer to a season-specific accident
than to a stable club property.

\paragraph{Penalties.}
The penalty shooting effect $w_{ts}$ of \eqref{eq:lp-w} is the clearest case
of a component that matters for outcomes while carrying no club-level signal,
and the two facts are worth separating. Penalties matter. A team receives
$1.39$ per match on average. The two sides receive an equal number in only
$26.6$ per cent of matches, and removing penalty goals from the record would
change the winner in $12.2$ per cent of them. The award rate more than doubled
over the window, from $0.96$ to $2.01$ per team-match, the largest
proportional change we measure. Conversion, however, is close to a constant.
Observed penalty conversion varies across team-seasons with a standard
deviation of $0.087$, while binomial variation alone, at the observed median of
$34$ attempts per team-season and a rate of $0.80$, would produce $0.069$.
Almost the whole spread between clubs is sampling noise. The model reaches the
same conclusion from the other direction. In the configuration that retains
$w_{ts}$, its persistence is $0.32$ with an interval of $[-0.11, 0.77]$,
spanning almost the whole prior. Removing it costs nothing: the difference in
expected log predictive density is $2.9$ with a standard error of $4.1$.

We therefore keep the penalty component and drop its team effect. Two cautions
apply: officiating varies between matches and the model does not separate that
variation from the shooter, and the opposing goalkeeper enters only through the
season intercept once the team effect is gone. Neither changes the finding that
a club-level shooting ability cannot be identified from one or two penalties
per match. Folding penalties into the action component instead would leave
every match total unchanged and cost nothing in prediction, which is part of
why the totals models compete well season-ahead, but it would discard the two
findings above: that penalty awards are the fastest-moving feature of the
league, and that penalty conversion, unlike man-up conversion, is not a
property of the club.

\paragraph{Correlations between components.}
The ten largest posterior correlations between ability dimensions are all
positive, between $0.37$ and $0.60$, and three of them carry the substance.
Defensive even-strength quality and the suppression of opponents' man-up
conversion correlate at $0.60$ $[0.37, 0.79]$: clubs that defend well at even
strength also defend well a player short. Attacking quality and man-up
conversion correlate at $0.45$ $[0.18, 0.68]$: clubs that score well score well
in both phases. Attack and defence correlate at $0.42$ $[0.19, 0.63]$: the good
clubs are good at both, which is the ordering the league table records.

\begin{figure}[t]\centering
\includegraphics[width=\textwidth]{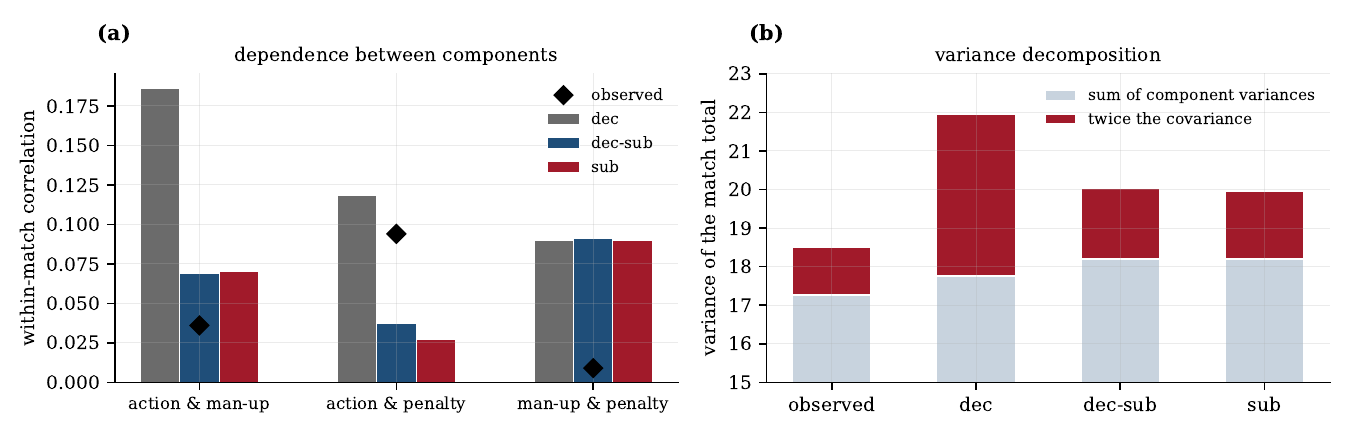}
\caption{Posterior predictive dependence between components under three
specifications: \textsc{dec} without a substitution term, and \textsc{dec-sub}
and \textsc{sub} with it. Panel (a) gives the within-match correlation of
component goal counts, observed and replicated; panel (b) decomposes the
variance of the match total into the sum of component variances and twice
their covariance. The substitution term removes most of the excess covariance
that \textsc{dec} manufactures.}
\label{fig:ppc}
\alttext{Panel (a) compares the within-match correlation between pairs of scoring components, observed against replicated, for three configurations. The configuration without a substitution term replicates far more correlation between action and man-up goals than is observed, while the two with the term are close to the observed value. Panel (b) decomposes the variance of the match total into the sum of component variances and twice their covariance, showing that the excess variance comes from the covariance term.}
\end{figure}

\paragraph{Posterior predictive checks.}
Figure~\ref{fig:ppc} reports the checks that motivated the substitution term. Under the decomposed model without it, each component
is individually well calibrated but the match total is not: the replicated
variance-to-mean ratio is $1.981$ against an observed $1.671$, with a tail
probability of $1.000$. The diagnosis is in the dependence. Observed action
and man-up goals are almost uncorrelated within a match, at $0.036$, while the
model replicates $0.186$; decomposing the variance of the total shows an
observed covariance contribution of $0.63$ against a replicated $2.10$. The
correlated ability structure, which the data clearly support at the level of
seasons, produces a within-match co-movement that does not exist. Adding
the substitution term brings the replicated correlation to $0.070$ and the
covariance contribution to $0.88$, and the total's variance-to-mean ratio to
$1.800$.

Two residual misfits remain and we report them plainly. The correlation
between man-up and penalty goals is still over-stated, $0.090$ against an
observed $0.009$, because the substitution term enters only the action rate
and cannot break the link between the two opportunity processes. Man-up
goals also remain slightly over-dispersed, with a replicated variance-to-mean ratio
of $1.252$ against $1.125$. We can identify the source of the second. Conditional on the fitted rates,
man-up attempts are under-dispersed relative to the Poisson, with a Pearson
residual variance of $0.881$ $[0.863, 0.899]$. Penalties awarded behave
similarly, at $0.934$ $[0.898, 0.972]$, while action goals are exactly Poisson
at $1.004$ $[0.982, 1.028]$. Refereed opportunity counts are more regular than
a Poisson process allows. Both points are taken up in
Section~\ref{sec:discussion}.

\begin{figure}[t]\centering
\includegraphics[width=\textwidth]{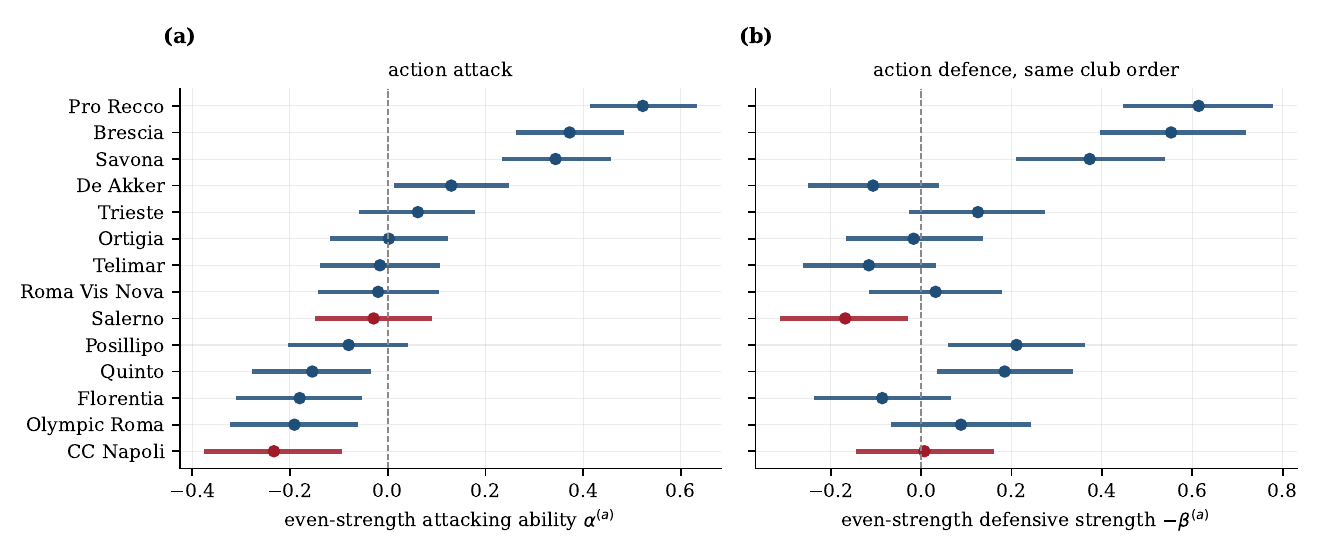}
\caption{Even-strength ability of the 2025--26 clubs under \textsc{sub},
posterior means with ninety per cent credible intervals. Panel (a): attacking
ability $\alpha^{(a)}$. Panel (b): defensive strength $-\beta^{(a)}$, positive
values indicating that opponents score less. Clubs are ordered by attack in
both panels; the two promoted clubs are highlighted.}
\label{fig:teams}
\alttext{Two horizontal interval plots of the fourteen clubs of the 2025--26 season, ordered by even-strength attacking ability. Panel (a) shows attack, with Pro Recco highest at about 0.52 and the two promoted clubs at the foot. Panel (b) shows defensive strength in the same club order; the top three clubs lead both panels, while Posillipo and Quinto defend well despite weak attack and Salerno has the weakest defence.}
\end{figure}

\paragraph{Team profiles.}
Figure~\ref{fig:teams} shows the 2025--26 clubs' even-strength attacking
ability $\alpha^{(a)}$ and defensive strength $-\beta^{(a)}$, the latter signed
so that a positive value means the opponent scores less, with clubs in the
same order in both panels. The league is wide. Pro Recco, at $0.52$ $[0.41,
0.63]$ on attack, scores at roughly twice the even-strength rate of the weakest
clubs, and it leads the defensive ordering as well at $0.61$ $[0.45, 0.78]$.
The top three are the top three on both panels, which is the $0.42$
correlation between attack and defence made visible. Below them the panels
diverge. Posillipo and Quinto, weak on attack, are the fourth and fifth best
defences; De Akker and Telimar show the opposite profile. The two entrants
differ too: Canottieri Napoli at $-0.23$ on attack defends at the league
average, while Salerno, at $-0.03$ on attack, has the weakest defence in the
league at $-0.17$ $[-0.31, -0.03]$, the only club whose defensive interval
excludes zero on the negative side.

\begin{figure}[t]\centering
\includegraphics[width=\textwidth]{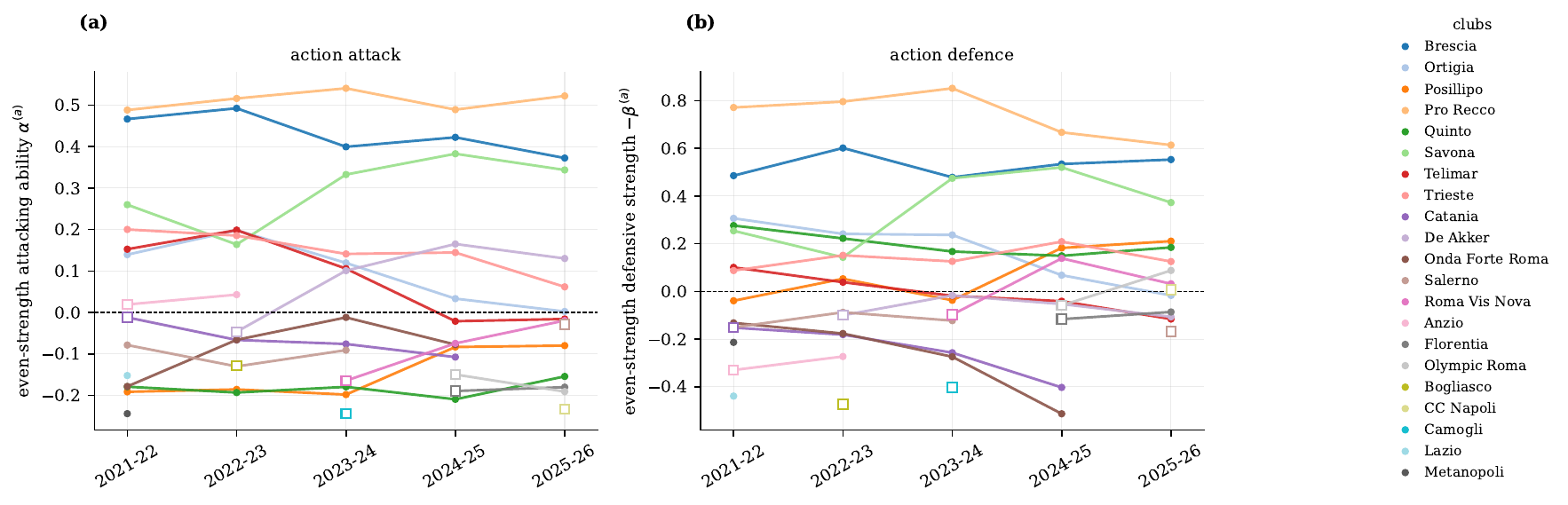}
\caption{Even-strength attacking ability $\alpha^{(a)}$ (a) and defensive
strength $-\beta^{(a)}$ (b) of every club across the five seasons, posterior
means under \textsc{sub}. Open squares mark a club's season of entry by
promotion, and a line is broken where a club was absent from the league.}
\label{fig:evolution}
\alttext{Two line charts tracing all twenty-one clubs' even-strength attacking and defensive ability across the five seasons, one colour per club. Pro Recco and Brescia stay at the top of both panels throughout, Savona rises steadily to third, and the promoted clubs, marked with open squares, enter at or below zero on both dimensions and mostly stay there.}
\end{figure}

The whole window is in Figure~\ref{fig:evolution}, which traces every club's
even-strength attack and defence across the five seasons. Two features stand
out. The top of the league is stable: Pro Recco and Brescia hold the first two
places on both dimensions in every season, and the one club to move through
the order, Savona, does so gradually, rising from the middle of the pack in
2021--22 to third on both axes by 2024--25. And the entrants, marked at their
season of entry, arrive at or below the league average on both dimensions in
every case. Most stay there. De Akker is the exception on attack, climbing to
a positive value from its second season, and Roma Vis Nova and Olympic Roma
reach an average defence while their attack does not recover.

\subsection{Predictions and the value of the decomposition}\label{sec:results-predictions}

Table~\ref{tab:scores} reports the out-of-sample scores on both tiers.

\begin{table}[t]\centering\small
\caption{Out-of-sample scores. Season-ahead: averages over the four
forward-chained folds, each predicting a whole season from the seasons before
it. Round-by-round: averages over seven blocks of the 2025--26 season,
refitting after every third round. Log score, CRPS and interval width refer to
total goals per team-match; RPS and accuracy to the match result. Best value
in each column of each block in bold; coverage is nominal 0.90.}
\label{tab:scores}
\begin{tabular}{lrrrrrr}\toprule
& log score & CRPS & RPS & accuracy & width & coverage \\\midrule
\multicolumn{7}{l}{\emph{season-ahead}} \\
\textsc{tot-s} & -2.6773 & 1.9886 & 0.1660 & 0.687 & \textbf{11.30} & 0.925 \\
\textsc{tot-d} & -2.7101 & 2.0451 & 0.1722 & 0.698 & 12.95 & 0.948 \\
\textsc{tot-dc} & -2.7048 & 2.0353 & 0.1735 & 0.694 & 12.84 & 0.947 \\
\textsc{dec} & -2.6935 & 2.0319 & 0.1729 & 0.700 & 12.46 & 0.945 \\
\textsc{dec-h} & -2.6912 & 2.0295 & 0.1725 & 0.701 & 12.44 & 0.944 \\
\textsc{dec-p} & -2.6936 & 2.0286 & 0.1728 & 0.701 & 12.51 & 0.946 \\
\textsc{dec-sub} & -2.6874 & 2.0279 & 0.1720 & 0.698 & 11.74 & 0.928 \\
\textsc{sub} & -2.6868 & 2.0268 & 0.1722 & 0.694 & 11.77 & 0.928 \\
\textsc{sub-s1} & \textbf{-2.6755} & \textbf{1.9828} & \textbf{0.1636} & 0.723 & 12.70 & 0.945 \\
\textsc{sub-s2} & -2.7477 & 2.3963 & 0.1676 & \textbf{0.730} & 25.39 & 0.948 \\
\textsc{sub-s3} & -2.7037 & 2.0949 & 0.1643 & 0.726 & 15.66 & 0.946 \\
\addlinespace
\multicolumn{7}{l}{\emph{round-by-round}} \\
\textsc{tot-s} & -2.6158 & 1.8325 & 0.1592 & \textbf{0.738} & 12.01 & 0.961 \\
\textsc{dec} & -2.6177 & 1.8135 & 0.1382 & 0.731 & 12.31 & 0.968 \\
\textsc{sub} & \textbf{-2.5958} & \textbf{1.7891} & \textbf{0.1372} & \textbf{0.738} & \textbf{11.26} & 0.951 \\
\textsc{sub-s1} & -2.5987 & 1.8024 & 0.1393 & 0.731 & 11.27 & 0.951 \\
\bottomrule\end{tabular}\end{table}

\paragraph{Season-ahead prediction.}
Over the four forward-chained folds, no configuration separates from any other
on the total-goals log score. The simplest model in the set, a static double
Poisson on match totals, attains the best mean log score at $-2.677$ against
\textsc{sub}'s $-2.687$, and every paired difference lies within two standard
errors of zero. We state this negative result plainly. When a whole season must be predicted
before it starts, most of the uncertainty lies in the league-wide scoring
level, which moved by thirty-six per cent over our window. Team-level structure
does not reduce that uncertainty.

\paragraph{Round-by-round prediction.}
Once the scoring level is known and the clubs' states are informed by data,
the picture reverses. Over seven round-blocks of the final season \textsc{sub}
attains the best score on every measure, with the sharpest
intervals at $11.26$ goals and the best-calibrated coverage at $0.951$.
Against the decomposed model without the substitution term, the difference in
log score is $0.0219$ with a standard error of $0.0047$, more than four
standard errors. The totals model's advantage disappears. The differences are small on the
scale of a single match, a few hundredths of a nat per team-match, but they
accumulate over a season of twenty-six rounds and, as
Sections~\ref{sec:results-postseason} and \ref{sec:results-midseason} show,
they are large enough to change which club a forecast names for relegation.
The contrast between the two tiers is itself a result. The decomposition improves accuracy when the
league level is known, and it aids interpretation throughout.

\paragraph{The entrant prior.}
Table~\ref{tab:entrant} reports the entrant prior under the three strategies. The estimates are stable across strategies and
economically large. Under S1 an entrant begins at $-0.302$ $[-0.473, -0.132]$
on even-strength defence, $-0.277$ $[-0.405, -0.145]$ on man-up conversion,
$-0.203$ $[-0.323, -0.085]$ on conceding penalties and $-0.178$ $[-0.303,
-0.049]$ on even-strength attack. All four intervals exclude zero, and the
corresponding estimates under S2 and S3 differ by less than a fifth of an
interval width.

\begin{table}[t]\centering\small
\caption{Entrant prior means under three promotion strategies. Posterior means
with 90 per cent credible intervals. Negative values indicate that promoted
clubs begin below the league average on that dimension.}
\label{tab:entrant}
\begin{tabular}{lccc}\toprule
dimension & S1 & S2 & S3 \\\midrule
$\alpha^{(a)}$ action attack & -0.178 [-0.30, -0.05] & -0.176 [-0.32, -0.03] & -0.197 [-0.32, -0.07] \\
$\beta^{(a)}$ action defence & -0.302 [-0.47, -0.13] & -0.318 [-0.46, -0.17] & -0.295 [-0.48, -0.11] \\
$\alpha^{(x)}$ man-up drawn & +0.029 [-0.04, +0.10] & +0.019 [-0.05, +0.09] & +0.017 [-0.05, +0.08] \\
$\beta^{(x)}$ man-up conceded & -0.036 [-0.13, +0.06] & -0.043 [-0.13, +0.04] & -0.045 [-0.15, +0.06] \\
$c$ man-up conversion & -0.277 [-0.41, -0.14] & -0.276 [-0.41, -0.14] & -0.269 [-0.40, -0.14] \\
$k$ man-up suppression & -0.100 [-0.26, +0.06] & -0.111 [-0.29, +0.07] & -0.121 [-0.28, +0.04] \\
$\alpha^{(p)}$ penalty drawn & -0.226 [-0.46, -0.01] & -0.236 [-0.49, -0.00] & -0.244 [-0.51, +0.02] \\
$\beta^{(p)}$ penalty conceded & -0.203 [-0.32, -0.09] & -0.214 [-0.34, -0.09] & -0.214 [-0.34, -0.08] \\
\bottomrule\end{tabular}\end{table}

One dimension is the exception, and it is the most informative part of the
result. Entrants are not deficient at drawing exclusions, $+0.029$ $[-0.039,
0.099]$, the only dimension whose posterior mean is positive under all three
strategies. Promoted clubs earn the extra man at the league rate from their
first match. They cannot convert it, and they cannot defend, either at even
strength or a player short. Read against the register, this says precisely
what transfers between divisions. The entrants converted $41.8$ per cent of
their second-division man-up situations against an A1 norm of $37.9$, but that
advantage belonged to the opposition they faced, not to the club. Their ability
to draw fouls survives the change of level.

\paragraph{The second-division covariates.}
The entrant prior earns its place out of sample: in season-ahead prediction
the configuration carrying it attains the best ranked probability score in the
set, $0.1636$ against $0.1722$ without it, and its advantage in continuous
ranked probability score, $0.0440$ with a standard error of $0.0179$, is the
only comparison in the tier that clears two standard errors. The covariates
do not. Every loading in $\Lambda$, for the promotion record and for the
carry-over block alike, has a posterior interval that spans zero by a wide
margin, and out of sample they do harm. The record covariates raise the mean
absolute error from $2.82$ to $8.19$ goals per team-match, because in the
first fold one training season and two entrants leave the loadings at their
prior and the implied entrant means swing by several units on the log scale.
The carry-over strategy fails in the one fold where it is active, 2025--26,
Salerno's return season, with the worst log score of the four at $-3.034$.

Knowing that a club has just been promoted is therefore useful. Knowing how it
was promoted, by margin, position or group, is not, and nor is a previous
spell in the top flight, at least with the single episode our window contains.
\citet{ridall2025} report that lower-division form is barely informative for
promoted English football clubs, and our results agree, with a mechanism the
decomposition supplies: promoted clubs are not uniformly weaker by an amount
the record could reveal, but the components in which they excelled below are
the ones that fail to transfer.

\subsection{Predicting the post-season}\label{sec:results-postseason}

The quantities of substantive interest in this league are decided after the
regular season, and they are decided by short knockout series. We therefore
repeat the exercise a governing body or a club would face on the evening the
regular season ends. For each season the model is fitted to every earlier
season in full and to the target season's regular season only, and the
post-season bracket is then simulated forward from the seeding that the
regular season produced. Nothing observed after the regular season enters the
fit. Series are best of three with the higher seed hosting the opening match,
following the observed schedule, and regulation ties are resolved by a fair
shoot-out, which the data record as a separate mechanism and which the model
does not attempt to predict.

Two clubs are relegated each season. The club finishing last in the regular
season goes down directly, so its fate is settled before the bracket begins
and we treat it as known rather than predicted. The second place is decided by
the play-out, a loser bracket in which the two semifinal losers meet and the
loser of that final is relegated. This rule reproduces all ten realised
relegations of the window exactly.

\paragraph{The championship.} Pro Recco won all five titles. The model, knowing
only the regular seasons, gave it $0.88$, $0.78$, $0.95$, $0.60$ and $0.82$,
for a mean log score of $-0.235$ and a Brier score of $0.025$. Brescia took
almost all of the remainder, at $0.11$, $0.22$, $0.03$, $0.32$ and $0.17$, and
Savona was the only other club to reach as much as $0.07$, in 2024--25. That
was the season in which the model hedged most, and the one in which Brescia
took the final to a third match.

\begin{figure}[t]\centering
\includegraphics[width=0.6\textwidth]{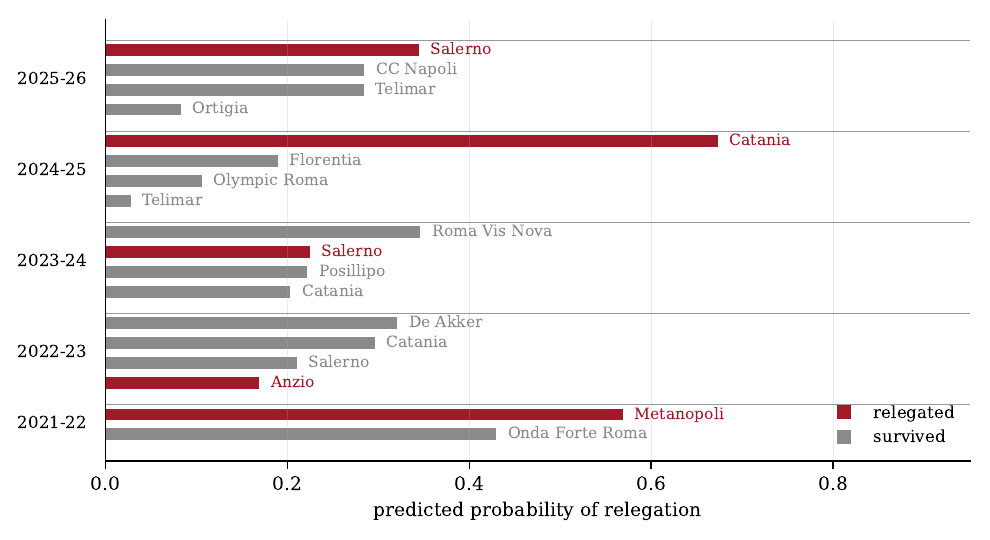}
\caption{Relegation forecasts made on the evening each regular season ended,
before any play-out match was played: the predicted probability for each of
the eighteen play-out participants, grouped by season, with the clubs
subsequently relegated shown in colour.}
\label{fig:releg}
\alttext{A horizontal bar chart of the predicted relegation probability for each of the eighteen play-out participants, grouped by season, with the clubs subsequently relegated highlighted. The relegated club is the highest-probability pick in three of the five seasons, second of four in one, and last of four in 2022--23.}
\end{figure}

\paragraph{The play-out.}
Across the eighteen play-out participants of the five seasons the model attains
a Brier score of $0.159$ against $0.194$ for a forecast giving every club in
the play-out the same probability, and a mean log score of $-0.491$ against $-0.577$
(Figure~\ref{fig:releg}). The improvement is real but modest, and the ordering
tells the story more clearly than the scores. The relegated club was the
model's most likely candidate in three of the five seasons, second of four in
another, and last of four in 2022--23, when Anzio went down having been given
$0.170$.

This describes the competition rather than a failure of the model. The clubs
that reach a play-out are those the league could not separate over the regular
season. The ability differences that make Pro Recco a $0.9$ favourite for the
title do not exist among them. Probabilities between $0.17$ and $0.35$ across a
four-club field are therefore the right answer. One play-out is different. In
2024--25 the model gave Catania $0.674$, its only confident call, and that was
the season in which the regular season had separated a club. Catania went
down.

The exercise also gives the promoted-club priors a concrete use. Six of the
eighteen play-out participants were clubs promoted within the window, and in
2025--26 the model made Salerno, promoted that season, the most likely of four
to be relegated at $0.346$; Salerno lost the play-out final to Telimar and went
down. Section~\ref{sec:results-predictions} showed that promotion status
carries predictive information; the relegation bracket is where that
information is worth having.

\subsection{Forecasting the relegation zone from mid-season}\label{sec:results-midseason}

Section~\ref{sec:results-postseason} begins once the play-out field is known.
The decision that matters to a club, however, is taken earlier, when the field
is still open. We therefore repeat the exercise from the middle of the regular
season, fitting to every earlier season in full and to the target season's
first thirteen rounds, and simulating the remaining fixtures, which are known
in advance, to obtain the distribution of the final table.

Round thirteen is the natural cut throughout the window. In 2022--23,
2024--25 and 2025--26 the league plays a double round robin over twenty-six
rounds, so it is the half-way point; in 2021--22 and 2023--24 the league plays
a single round robin over thirteen rounds and then splits into an upper and a
lower group of seven, so it is the last round at which every club still faces
a common field. To trace how the picture resolves we repeat the forecast at
rounds five and nine as well. Two quantities are predicted: the identity of
the club that will finish last and be relegated directly, and the composition
of the play-out field, namely the four clubs finishing immediately above it.

Throughout, the natural benchmark is not an equal-probability forecast but the
league table itself, which anyone can read. We therefore report alongside each
forecast what the table alone would have said: that the club currently last
finishes last, and that the four clubs immediately above it contest the
play-out.

\paragraph{Direct relegation.}
Table~\ref{tab:midseason} reports the results.
By round nine the model identifies the directly relegated club as the most
probable in all five seasons, a standard the league table does not reach until
round thirteen; at round five the model is correct in four seasons against the
table's three. The probabilities are strong but not reckless. At round thirteen they run from
$0.740$ for Onda Forte Roma in 2024--25, the least clear of the five, to
$0.991$ for Camogli in 2023--24. The Brier score is $0.003$, against $0.066$
for giving each of the fourteen clubs the same probability. The club that will be relegated
directly is therefore settled by the end of the first third of the season, and
a model that pools information across seasons and components can say so a month
before the table can.

\begin{table}[t]\centering\small
\caption{Forecasting the relegation zone from mid-season, over the five
seasons and their eighteen eventual play-out participants. The league table
benchmark predicts that the club currently last finishes last and that the
four clubs immediately above it contest the play-out. A forecast giving every club the same probability scores $0.066$ and $0.192$
respectively.}
\label{tab:midseason}
\begin{tabular}{lrrrrrr}\toprule
& \multicolumn{3}{c}{club relegated directly} & \multicolumn{3}{c}{play-out field} \\
\cmidrule(lr){2-4}\cmidrule(lr){5-7}
forecast made at & Brier & log score & correct & Brier & log score & in top four \\\midrule
round 5  & 0.028 & $-0.088$ & 4 of 5 & 0.137 & $-0.413$ & 10 of 18 \\
round 9  & 0.008 & $-0.036$ & 5 of 5 & 0.111 & $-0.344$ & 14 of 18 \\
round 13 & 0.003 & $-0.016$ & 5 of 5 & 0.098 & $-0.282$ & 13 of 18 \\\addlinespace
league table, round 5  & & & 3 of 5 & & & 11 of 18 \\
league table, round 9  & & & 4 of 5 & & & 12 of 18 \\
league table, round 13 & & & 5 of 5 & & & 12 of 18 \\
\bottomrule\end{tabular}\end{table}

\paragraph{The play-out field.}
Predicting which four clubs will finish immediately above last is genuinely
uncertain. A handful of points separates them and a third of the season
remains. The model's Brier score falls from $0.137$ at round five to $0.098$ at
round thirteen, against $0.192$ for the equal-probability forecast. Forecasting at round
nine, its four most likely clubs contain fourteen of the eighteen eventual
play-out participants, against twelve for the league table. The improvement is
small but consistent, and it comes at a point when four of those participants
were not in the table's bottom four at all.

\paragraph{The three stages together.} Taken with
Section~\ref{sec:results-postseason}, the relegation decision resolves in
three stages of sharply different predictability. The reference at each stage is a forecast that assigns
every eligible club the same probability, one in fourteen of finishing last,
four in fourteen of reaching the play-out and one in four of losing it, which
is what a forecaster with no information about the clubs would say. Lower
Brier scores are better, and the model beats that reference at every stage,
but by shrinking margins. Which club is relegated directly is all but settled
by round nine: $0.008$ for the model against $0.066$ for the reference, an
eightfold improvement. Which clubs join it in the play-out is partly
predictable: $0.111$ against $0.192$. Which of those clubs is then relegated
is close to a lottery: $0.159$ against $0.194$, because the play-out field
consists precisely of the clubs the regular season could not separate.
The same model, applied at three points in one competition, is in turn nearly
certain, moderately informative, and barely better than the reference. This
progression reflects the competition format, not the model. A governing body
deciding how much of the relegation decision to delegate to a short knockout
needs exactly this quantity.

\section{Discussion}\label{sec:discussion}

Two of our results are invisible to a model of match totals. Home advantage in
water polo is a special-situation effect: largest in man-up conversion, half
that size in even-strength scoring and in the drawing of exclusions, and
absent from penalties in both their award and their conversion. The home
effect appears where execution under structure matters and disappears at the
two moments most isolated from crowd and context, and constraining the
components to share one home advantage, as a totals model implicitly does,
destroys the finding. The other is the anatomy of promotion. Entrants arrive
materially below the league on even-strength defence, man-up conversion and
penalty discipline, but at the league rate on drawing exclusions; their
second-division man-up conversion advantage does not survive the change of
division, whereas their foul-drawing does. This sharpens the negative result of
\citet{ridall2025}. Lower-division form is uninformative not because promoted
clubs are indistinguishable from one another but because the record aggregates
components that transfer with components that do not, and our covariate-based
strategies fail out of sample, in short training windows actively degrading
forecasts, which is a caution against the intuition that more prior
information must help.

The relegation results support the paper's motivating claim, and a single
accuracy figure would have hidden them. Sections~\ref{sec:results-postseason}
and \ref{sec:results-midseason} show the decision resolving in three stages:
round nine settles the club relegated directly, the composition of the
play-out field is partly predictable, and the play-out itself is close to a
lottery among four clubs the regular season could not separate. That the same
model is in turn nearly certain, moderately informative and barely better than
an equal-probability forecast reflects the format rather than the method. The
same asymmetry shows why relegation carries the outcome uncertainty of this
league: we predict the championship to a Brier score of $0.025$ and the
play-out only to $0.159$, against $0.194$ for equal probabilities. A governing
body weighing a change of format does not need to know how well a season can
be forecast on average. It needs to know how much of the relegation decision
is being delegated to a three-match series between near-equal clubs.

Two features of the evaluation carry lessons beyond this application. When a
whole season must be forecast in advance, the decomposed model cannot be
distinguished from a static double Poisson, because the season-ahead problem
is dominated by uncertainty in the league-wide scoring level and team-level
structure cannot reduce it; when the season is under way, the decomposition is
clearly better. Reporting one tier alone would have supported either an
over-claim or an under-claim. The playing rules also changed within our
window. The season-indexed intercepts absorb the shift, so ability estimates
remain comparable, but the league of 2025--26 is not the league of 2021--22,
and a forecast made before a rule change will be worse than our season-ahead
results suggest, a general caution for league models fitted across regulatory
reform.

Several limitations bound these conclusions. Exposure time for man-up
situations is not recorded in the official reports, so the model counts
opportunities rather than measuring the seconds a team spends with the
advantage; timestamped exclusions would allow a rate model. The analysis is at
team level, and the man-up conversion effects in particular aggregate a
shooting structure plausibly attributable to a few players. Abilities are
constant within a season, a simplification that the round-by-round tier
compensates by refitting but does not remove. Regulation draws are modelled as
draws and the shoot-out is not predicted, which matters most in the play-out
series, where a single shoot-out can decide a relegation. The promotion
register spans ten episodes, a small sample for estimating a transfer matrix,
and the carry-over strategy is identified from a single returning club, so its
failure is an absence of evidence rather than evidence of absence. Our own
diagnostics point to two refinements: refereed opportunity counts are
under-dispersed relative to the Poisson, at conditional dispersion $0.88$ for
man-up attempts and $0.93$ for penalties, which a Conway--Maxwell--Poisson
observation model would accommodate; and the substitution term enters only the
action rate, leaving the dependence between the two opportunity processes
over-stated, so a joint allocation of possessions across all three components
is the natural generalisation.

The framework transfers directly. The women's Serie A1 is reported in the
identical format and would allow a comparison of component structure and home
advantage across the two competitions under one protocol, and other national
leagues and the Champions League differ in format rather than in reporting.
More broadly, the decomposition with a substitution term is not specific to
water polo: any sport in which possessions terminate in mutually exclusive
ways, and in which the resulting opportunities are separately recorded, admits
the same treatment.

\section*{Conflicts of interest}
The author declares no competing interests.

\section*{Funding}
This work was supported by the Spanish State Research Agency (AEI) through
the project SPHERES, grant PID2023-153222OB-I00.

\section*{Data availability}
The dataset assembled for this article, comprising the 940 Serie A1 match
records with full component detail and the second-division promotion register,
together with the Stan and R code and the scripts that reproduce every table
and figure, are available at \url{https://github.com/manueleleonelli/waterpolo}.

\section*{Acknowledgements}
During the preparation of this work the author used generative AI tools to
support the coding and writing. After using these tools, the author reviewed
and edited the content as needed and takes full responsibility for the content
of the published article.

\bibliographystyle{plainnat}
\bibliography{reference}

@article{maher1982, author={Maher, M. J.}, title={Modelling association football scores},
 journal={Statistica Neerlandica}, volume={36}, number={3}, pages={109--118}, year={1982}}

@article{dixon1997, author={Dixon, M. J. and Coles, S. G.},
 title={Modelling association football scores and inefficiencies in the football betting market},
 journal={Journal of the Royal Statistical Society: Series C}, volume={46}, number={2}, pages={265--280}, year={1997}}

@article{karlis2003, author={Karlis, D. and Ntzoufras, I.},
 title={Analysis of sports data by using bivariate {P}oisson models},
 journal={The Statistician}, volume={52}, number={3}, pages={381--393}, year={2003}}

@article{baio2010, author={Baio, G. and Blangiardo, M.},
 title={Bayesian hierarchical model for the prediction of football results},
 journal={Journal of Applied Statistics}, volume={37}, number={2}, pages={253--264}, year={2010}}

@article{glickman1999, author={Glickman, M. E.}, title={Parameter estimation in large dynamic paired comparison experiments},
 journal={Journal of the Royal Statistical Society: Series C}, volume={48}, number={3}, pages={377--394}, year={1999}}

@article{koopman2015, author={Koopman, S. J. and Lit, R.},
 title={A dynamic bivariate {P}oisson model for analysing and forecasting match results in the {E}nglish {P}remier {L}eague},
 journal={Journal of the Royal Statistical Society: Series A}, volume={178}, number={1}, pages={167--186}, year={2015}}

@article{cattelan2013, author={Cattelan, M. and Varin, C. and Firth, D.},
 title={Dynamic {B}radley--{T}erry modelling of sports tournaments},
 journal={Journal of the Royal Statistical Society: Series C}, volume={62}, number={1}, pages={135--150}, year={2013}}

@article{titman2015, author={Titman, A. C. and Costain, D. A. and Ridall, P. G. and Gregory, K.},
 title={Joint modelling of goals and bookings in association football},
 journal={Journal of the Royal Statistical Society: Series A}, volume={178}, number={3}, pages={659--683}, year={2015}}

@article{ridall2025, author={Ridall, P. G. and Titman, A. C.},
 title={Modelling promoted teams in association football}, journal={Journal of Applied Statistics}, year={2025}, note={to appear}}

@book{ntzoufras2009, author={Ntzoufras, I.}, title={Bayesian Modeling Using {WinBUGS}}, publisher={Wiley}, year={2009}}

@article{lkj2009, author={Lewandowski, D. and Kurowicka, D. and Joe, H.},
 title={Generating random correlation matrices based on vines and extended onion method},
 journal={Journal of Multivariate Analysis}, volume={100}, number={9}, pages={1989--2001}, year={2009}}

@article{vehtari2017, author={Vehtari, A. and Gelman, A. and Gabry, J.},
 title={Practical {B}ayesian model evaluation using leave-one-out cross-validation and {WAIC}},
 journal={Statistics and Computing}, volume={27}, number={5}, pages={1413--1432}, year={2017}}

@article{stan2017, author={Carpenter, B. and Gelman, A. and Hoffman, M. D. and Lee, D. and Goodrich, B.
 and Betancourt, M. and Brubaker, M. and Guo, J. and Li, P. and Riddell, A.},
 title={{S}tan: a probabilistic programming language}, journal={Journal of Statistical Software},
 volume={76}, number={1}, pages={1--32}, year={2017}}

@article{glickman1998, author={Glickman, M. E. and Stern, H. S.},
 title={A state-space model for national football league scores},
 journal={Journal of the American Statistical Association}, volume={93}, number={441}, pages={25--35}, year={1998}}

@article{egidi2020, author={Egidi, L. and Ntzoufras, I.},
 title={A {B}ayesian quest for finding a unified model for predicting volleyball games},
 journal={Journal of the Royal Statistical Society Series C: Applied Statistics},
 volume={69}, number={5}, pages={1307--1336}, year={2020}}

@article{gabrio2021, author={Gabrio, A.},
 title={{B}ayesian hierarchical models for the prediction of volleyball results},
 journal={Journal of Applied Statistics}, volume={48}, number={2}, pages={301--321}, year={2021}}

@article{leonelli2026exceptional,
 title={How exceptional was the {B}ig {T}hree era? {E}xtremes and persistence in men's professional tennis},
 author={Leonelli, Manuele}, journal={arXiv preprint arXiv:2608.27362}, year={2026}}

@article{leonelli2026predicting,
 title={Predicting and understanding shooting performance in professional biathlon:
 a {B}ayesian approach},
 author={Leonelli, Manuele},
 journal={International Journal of Performance Analysis in Sport},
 volume={26}, number={2}, pages={366--385}, year={2026}, publisher={Taylor \& Francis}}

@article{fioravanti2023, author={Fioravanti, Federico and Delbianco, Fernando and Tohm\'e, Fernando},
 title={The relative importance of ability, luck and motivation in team sports: a {B}ayesian model of performance in the {E}nglish {R}ugby {P}remiership},
 journal={Statistical Methods \& Applications}, volume={32}, number={3}, pages={715--731}, year={2023}}

@article{ingram2019, author={Ingram, Martin},
 title={A point-based {B}ayesian hierarchical model to predict the outcome of tennis matches},
 journal={Journal of Quantitative Analysis in Sports}, volume={15}, number={4},
 pages={313--325}, year={2019}}

@article{lupo2010, author={Lupo, Corrado and Tessitore, Antonio and Minganti, Carlo and Capranica, Laura},
 title={Notational analysis of elite and sub-elite water polo matches},
 journal={Journal of Strength and Conditioning Research}, volume={24}, number={1},
 pages={223--229}, year={2010}}

@article{escalante2011, author={Escalante, Yolanda and Saavedra, Jose M. and Mansilla, Mirella and Tella, Victor},
 title={Discriminatory power of water polo game-related statistics at the 2008 {O}lympic {G}ames},
 journal={Journal of Sports Sciences}, volume={29}, number={3}, pages={291--298}, year={2011}}

@article{lupo2025, author={Lupo, Corrado and Li Volsi, Damiano and Brustio, Paolo Riccardo and Ungureanu, Alexandru Nicolae},
 title={Water polo coaches believe they gain an advantage by calling time-out before playing power-play, but is that really true?},
 journal={Frontiers in Psychology}, volume={16}, pages={1548905}, year={2025}}

\appendix

\section{The competition over the study window}\label{app:formats}

Three seasons, 2022--23, 2024--25 and 2025--26, followed the
conventional structure of a home-and-away round robin of twenty-six rounds. The 2021--22 season was scheduled the same way, but a wave of Covid-19 cases
suspended play in January 2022 and the federation replaced the entire return
round: from February the clubs were divided, by the standings after the
thirteen-round first leg, into a championship pool of seven and a relegation
pool of seven, each playing a further single round robin of six matches with
one club resting per round and first-phase points carried forward. The
2023--24 season then adopted the same split-phase structure by design, its
calendar pausing from mid December to late February around the winter
European and World Championships.

The post-season was stable in outline. The title went to best-of-three
play-off series among the top four, preceded in 2025--26 by two-legged
quarter-finals that widened the bracket to the top eight, and concluded that
season by a title final extended to best of five; positions five to eight
were settled by a parallel bracket of semifinals and placement finals, drawing
in the split seasons on the fifth to seventh of the championship pool together
with the winner of the relegation pool, and no bronze final was played in
2025--26. Relegation combined a direct spot for the bottom club with a
play-out bracket among the four clubs above it, the losers of the two
play-out semifinals meeting in a final whose loser joined the bottom club in
Serie A2; in the split seasons the bracket was seeded from the relegation
pool, and in 2021--22 it reduced to a single final between the pool's fifth
and sixth, the seventh going down directly. Throughout, series place the odd
games with the better-classified club.

Points were awarded as three for a win and one for a draw, with drawn matches
standing, until 2024--25. From 2025--26 no league match may end level: a
penalty shoot-out follows every regulation draw, with three points for a
regulation win, two for a shoot-out win, one for a shoot-out loss and none
for a regulation defeat. Level clubs are separated by the federation's
classification criteria, beginning with head-to-head records. These format
specifics are not a nuisance to be conditioned away: they change the mapping
from latent team strength to relegation risk, and the model of
Section~\ref{sec:model} is deliberately specified at the level of match
scoring, so that any format can be layered on top by simulation, as the
evaluation of Section~\ref{sec:promoted-eval} requires.

Twenty-one clubs appear, eight of them ever-present, and two enter by
promotion every year; all ten promotion episodes in the window came through
the Serie A2 play-offs. The pattern at the top could hardly be starker: Pro
Recco won all five titles, and every regular season except 2024--25, when the
club finished level with Brescia and the head-to-head rule, Brescia having
won the last-round meeting by a single goal, handed Brescia first place and
with it the home advantage in the deciding games of a final it nevertheless
lost. The bottom, by contrast, changed hands every year: sixteen of the
twenty-one clubs occupied a direct relegation place or contested a play-out
bracket at least once, and in three of the last four seasons a newly promoted
club went straight back down, Bogliasco in 2022--23, Camogli in 2023--24 and
Salerno in 2025--26.

\section{Data sources, extraction and validation}\label{app:data}

The match centre publishes, for every fixture, a structured page: the header
with the final and per-period scores, the two rosters with individual goal
tallies, a typed event feed labelling every goal by type, and a closing
summary sentence giving each side's man-up conversions over attempts and its
penalties awarded, with missed and saved penalties itemised. We scraped the
pages in August 2026, after the 2025--26 season had concluded, so every fixture
is a completed match and no result was revised after extraction. Published
sources for this sport report final scores only, and the component detail used
here has not been assembled before.

Extraction followed
a fixed protocol: goal-type counts (the numerators) are taken from the typed
event feed, while opportunity counts (the denominators, namely man-up attempts
and penalties awarded) are taken from the summary sentence, since failed
special situations do not generate feed entries; team identity in the summary
sentence is assigned by position rather than by name, because the sentence is
hand-typed and occasionally misnames a club; and action goals are obtained by
subtracting man-up and penalty goals from the team's regulation total.
The typed event feed is available in all five seasons. Sporadic truncations and
missing summary sentences are confined to the records catalogued below. We
scraped the data in August 2026, after the 2025--26 season had concluded, so
every fixture in the window is a completed match and no result was revised
after extraction.

Every extracted match is subjected to deterministic consistency checks: the
goal-type decomposition must reproduce the final score; converted penalties
cannot exceed penalties awarded, and awarded penalties must equal the sum of
scored, missed and saved; man-up conversions cannot exceed attempts; per-period
scores must sum to the regulation total, with shoot-out goals identified and
excluded; and roster goal tallies must sum to the team total. At round level,
each round must contain every club exactly once with a consistent home and away
mirror, and at season level the aggregated results are reconciled against the
official final standings. Matches failing any check enter a manual audit queue
and are resolved against the source page; across the 940 matches of the
five seasons, 105 records required manual correction, the most common faults
being transcription slips in the per-period scores of the official pages
themselves. Twenty-six team-match records retain at least one missing
component field, ten of them lacking only the man-up attempt count; the
corresponding likelihood terms are dropped at estimation. One match, awarded
five--nil after a forfeit, is retained for standings and excluded from every
scoring likelihood.

Shoot-outs are rare and concentrated in the final season. We record the winner
of each but model the regulation score, so a match level at the end of
regulation enters the likelihood as a draw and contributes a point to each
side. Predicting the shoot-out itself would mean estimating what is close to a
coin toss from a handful of observations, and we prefer to leave it out of the
model.

The promotion register that Section~\ref{sec:promoted} requires is complete:
all ten promotion episodes are recorded, together with the eight Serie A2
group tables in which they occurred and the $243$ second-division match records
of the entrants themselves, each reconciled against the published
classification.

\section{Model specification}\label{app:model}

Table~\ref{tab:priors} collects the priors used in every specification we
report. All hierarchical layers use a non-centred parametrisation, so the
standard normal variates in the table are the sampled quantities and the
scales multiply them; the league intercepts are anchored at the pooled
observed rates, which keeps the sampler away from the extreme rates an
unanchored prior on the log scale permits; and the persistence parameters
receive a normal prior truncated to $(-1, 1)$, favouring moderate to high
persistence without excluding a negative value. Clubs already in the league in
the first season draw their abilities from the stationary distribution implied
by $\rho_d$ and $\sigma_d$, with scale $\sigma_d / \sqrt{1 - \rho_d^2}$, and
clubs promoted into it draw from the entrant prior of
Section~\ref{sec:promoted-strategies}.

\begin{table}[t]\centering\small
\caption{Prior distributions. $\Normal^+$ denotes a half-normal, and
$\Normal_{(-1,1)}$ a normal truncated to the stationary region. The anchors
$a_k$ are the pooled observed rates on the modelling scale: $\log 6.0$ for
action goals, $\log 10.7$ for man-up attempts, $\logit 0.377$ for man-up
conversion, $\log 1.40$ for penalties awarded and $\logit 0.809$ for penalty
conversion.}
\label{tab:priors}
\begin{tabular}{llll}
\toprule
quantity & symbol & prior & note \\
\midrule
league intercepts & $\mu^{(k)}_{s}$ & $\Normal(a_k, 1)$ & one per component and season \\
home advantages & $\eta^{(k)}$ & $\Normal(0, 1)$ & one per component \\
substitution coefficients & $\boldsymbol{\gamma}$ & $\Normal_2(\boldsymbol{0}, 0.5^2 I)$ & standardised covariates \\
innovation scales & $\sigma_d$ & $\Normal^+(0, 1)$ & one per ability dimension \\
ability correlations & $\Omega$ & $\mathrm{LKJ}(2)$ & Cholesky parametrisation \\
persistence & $\rho_d$ & $\Normal_{(-1,1)}(0.6, 0.35^2)$ & one per ability dimension \\
standardised abilities & $\boldsymbol{z}_{ts}$ & $\Normal(\boldsymbol{0}, I)$ & non-centred \\
\addlinespace
entrant prior mean & $\boldsymbol{m}_P$ & $\Normal(\boldsymbol{0}, I)$ & promoted clubs only \\
covariate loadings & $\Lambda$ & $\Normal(0, 1)$ & entrywise \\
entrant prior scale & $\sigma_{P,d}$ & $\Normal^+(0, 1)$ & promoted clubs only \\
\bottomrule
\end{tabular}
\end{table}

\end{document}